\documentclass[conference]{IEEEtran}
\IEEEoverridecommandlockouts
\usepackage{cite}
\usepackage{amsmath,amssymb,amsfonts}
\usepackage{graphicx}
\usepackage{textcomp}
\usepackage{xcolor}
\usepackage{amsmath, amssymb}
\usepackage{algorithm}
\usepackage{algpseudocode}
\usepackage{physics}
\usepackage{float} 
\usepackage{subcaption}
\usepackage{booktabs}
\usepackage{multirow}
\usepackage{wrapfig} 
\usepackage{array}    
\usepackage{booktabs} 
\usepackage{caption}  
\usepackage{adjustbox}
\usepackage{graphicx}
\usepackage{soul}
\usepackage{tikz}
\usepackage[caption=false, font=footnotesize]{subfig}
\usetikzlibrary{shapes.geometric, arrows.meta, positioning, fit, backgrounds}
\usetikzlibrary{calc}

\definecolor{navy}{HTML}{1A365D}
\definecolor{slate}{HTML}{2D3748}
\definecolor{crimson}{HTML}{9B2C2C}
\definecolor{ice}{HTML}{EDF2F7}
\definecolor{bordergray}{HTML}{CBD5E0}

\def\BibTeX{{\rm B\kern-.05em{\sc i\kern-.025em b}\kern-.08em
    T\kern-.1667em\lower.7ex\hbox{E}\kern-.125emX}}
\begin{document}
\bstctlcite{IEEEexample:BSTcontrol}

\title{Practical Zero-Trust for Mission-Critical Robotic Fleets via Hardware Attestation and Packet Timing Watermarking\thanks{This paper has been accepted for presentation and publication at the 2026 IEEE World Forum on Public Safety Technology (WF-PST).}}



\author{
  \IEEEauthorblockN{
    Ryne Gonzales\textsuperscript{3}, 
    Ethan Liesdyanto\textsuperscript{3}, 
    Rex Worley\textsuperscript{3}, 
    Michael Frederick\textsuperscript{3}, 
    Jaewon Kim\textsuperscript{2}, 
    and Eman Hammad\textsuperscript{1,3}
  }
  \IEEEauthorblockA{
    Texas A\&M University, College Station, TX, USA \\
    \textsuperscript{1}iSTAR Laboratory,  
    \textsuperscript{2}Global Cyber Research Institute (GCRI) \\
    \textsuperscript{3}Department of Engineering Technology and Industrial Distribution \\
    Emails: \{enyr29, ethanliesdyanto, pyrotexas, mfre7770, j1k, eman.hammad\}@tamu.edu
  }
}

\maketitle


\begin{abstract}

Autonomous unmanned vehicles are vital to tactical missions, mission-critical public-safety operations like search and rescue and disaster response. However, their reliance on open wireless links and standard Robot Operating System (ROS 2) middleware exposes a broad cyber-physical attack surface. A compromise of these systems can disrupt real-time control loops, leading to mission failure or asset loss in high-stakes environments. This paper presents and empirically evaluates a layered, context-aware cybersecurity framework enforcing Zero-Trust principles for a multi-node robotic fleet over Wi-Fi. The framework integrates an active hardware root of trust (TPM 2.0), centralized in-band and out-of-band SIEM telemetry monitoring (ELK Stack and Kismet), and a non-cryptographic Inter-Packet Delay (IPD) timing watermark. Evaluated on a live ROS 2 mobile testbed under multi-layer exploits (OSI Layers 2–5), results demonstrate that while volume-based filters isolate brute denial-of-service floods, tracking the statistical sample kurtosis ($K$) of the embedded IPD watermark exposes stealthy Man-in-the-Middle command injections with complete detection accuracy without payload overheads.
\end{abstract}


\section{Introduction}
The rapid advancement of Unmanned Vehicles (UxVs), encompassing both Unmanned Aerial Vehicles (UAVs) and Unmanned Ground Vehicles (UGVs), led to widespread adoption across diverse commercial, industrial, and tactical application domains. In critical deployments, such as search-and-rescue operations, disaster response, and hazardous environment inspections, these autonomous platforms rely fundamentally on wireless communication links to maintain loop closure for real-time navigation, telemetry streaming, and multi-agent coordination~\cite{alcorn2025situational, alcorn2026darrms}. However, this intrinsic dependence on open, shared wireless channels introduces an expansive and highly vulnerable attack surface~\cite{farraj2024physical, farraj2025reading}. Ensuring secure, deterministic, and continuous network operation within dynamic, unmanaged, or actively contested environments remains a major challenge hindering the resilient deployment of autonomous robotic fleets.

\begin{figure}[!htbp]
  \centering
  \includegraphics[width=0.75\columnwidth]{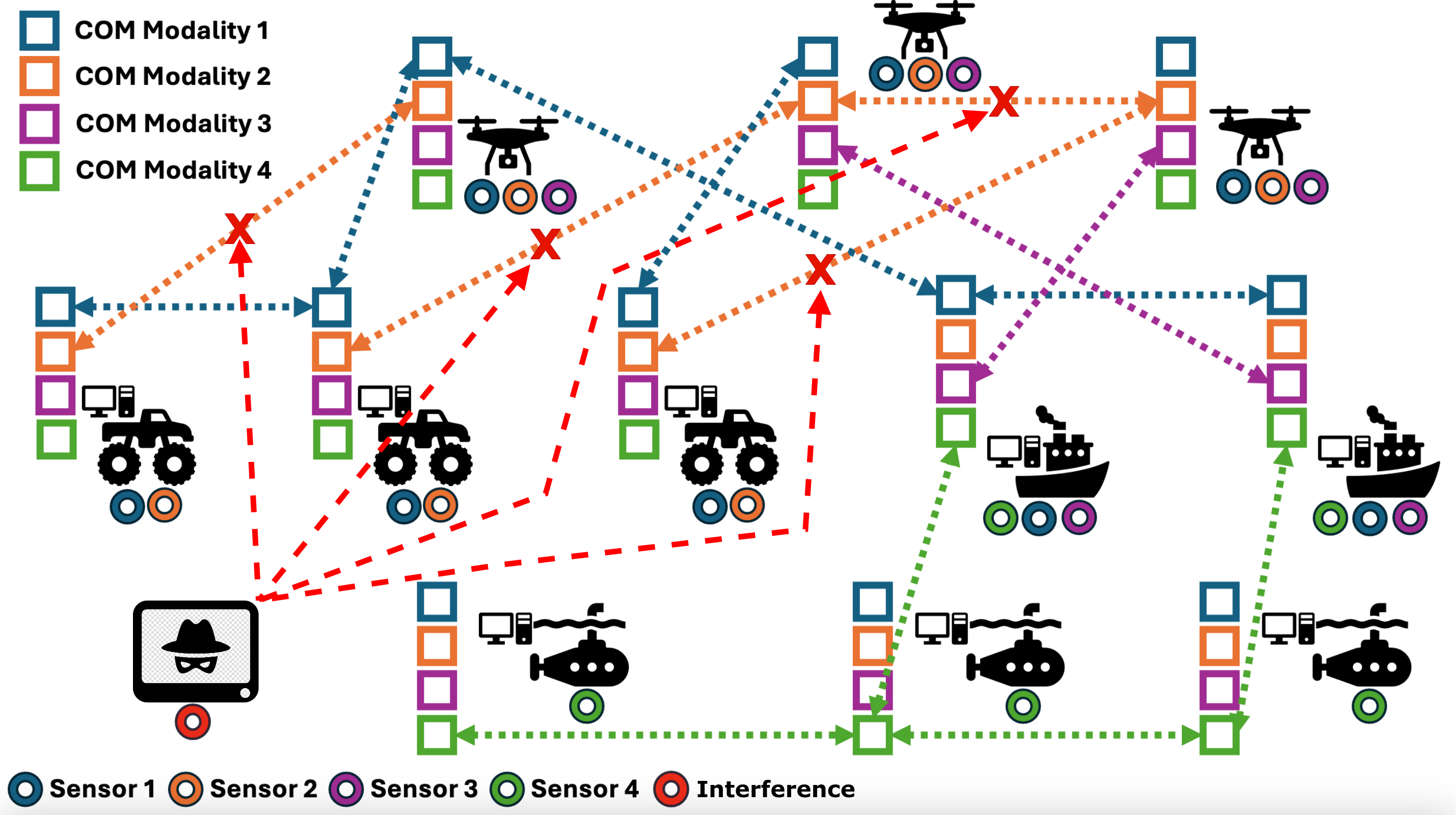}
  \caption{Example scenario illustrating potential interference attack targets in UxV communication.}
  \label{fig:overview}
  \vspace{-10pt}
\end{figure}

Real-world operations underscore these vulnerabilities: tactical electronic warfare like GPS jamming and RF spoofing induces control-loop disassociation and asset loss \cite{yaacoub2020security}, while control-link disruptions cause fleets to deviate from spatial boundaries or experience terminal failures. Wireless robotics must maintain bounded latency within non-stationary radio frequency (RF) environments where multi-path fading, thermal noise, and interference introduce stochastic variance into signal-to-noise ratios (SNR) and received signal strength indicators (RSSI). Interference escalates packet jitter and burst packet loss, negatively impacting feedback synchronization and state estimation. Because these metrics fluctuate during benign operations, rigid, static threshold-based anomaly detection models suffer from high false-alarm rates or fail to isolate low-power attacks. Reliably distinguishing environmental variability from malicious intervention requires multi-dimensional tracking of statistical features like latency distributions. Existing literature, however, often restricts its focus to theoretical link capacities or simulations, providing sparse empirical validation of how network anomalies translate to physical path deviation on live, hardware-constrained platforms \cite{yaacoub2020security, alrefaei2022survey}.

Simultaneously, threat vectors have evolved past primitive high-power broadband jamming toward intelligent, protocol-specific manipulation. Exploiting target communication protocols, timing distributions, and middleware formatting allows adversaries to execute selective packet dropping, localized de-authentication flooding, or man-in-the-middle (MITM) data injection using low-cost Software Defined Radios (SDRs). While broadband RF jamming presents a physical-layer threat, its spectral footprint exposes the adversary's presence \cite{yaacoub2020security, alrefaei2022survey, zeng2026gnss}. Consequently, modern tactical adversaries pivot toward stealthier exploits across OSI Layers 2 through 5. By exploiting the lack of native authentication in legacy 802.11 management frames and default Robot Operating System (ROS 2) Data Distribution Service (DDS) discovery middleware \cite{macenski2022robot, deng2022security, diluoffo2019credential}, attackers can achieve operational denial or command-hijacking without altering the ambient RF noise floor. 

This paper introduces a localized, context-aware cybersecurity framework integrating a hardware root of trust, centralized SIEM telemetry monitoring, and packet-level Inter-Packet Delay (IPD) watermarking to secure resource-constrained robotic networks. Beyond theoretical simulations, we provide live hardware testbed evaluations demonstrating how specific network exploits across OSI Layers 2–5 degrade real-time robotic state estimation and trajectory control loops as illustrated conceptually in Fig.~\ref{fig:overview}. Finally, the framework establishes a robust mechanism to isolate sophisticated, low-power on-path attacks, validating that while raw network jitter metrics fail to flag active MITM command injections, statistical kurtosis tracking of the embedded IPD watermark achieves complete detection accuracy.

\section{Background and Related Work}
Prior efforts in hardening autonomous mobile assets have concentrated primarily on the physical and link layers. Many works mitigate intentional RF broadband interference, cognitive jamming, and GPS spoofing through SDR spectrum monitoring, spatial nulling antennas, and adaptive SNR thresholds \cite{yaacoub2020security, alrefaei2022survey, zeng2026gnss, farraj2025reading, farraj2024physical}. These strategies operate below the data link layer and remain blind to protocol-specific exploits targeting network or middleware stacks. Software-centric research focuses on the application layer, demonstrating vulnerabilities within the ROS 2 ecosystem and evaluating security wrappers, mandatory access controls, or application-level payload authentication to prevent command tampering \cite{macenski2022robot, deng2022security, diluoffo2019credential}. 

Yet, application-centric defenses often ignore built-in mitigations for lower-layer denial-of-service (DoS) vulnerabilities that interrupt communication entirely. Furthermore, although network intrusion systems have been proposed using flow-volume threshold modeling \cite{rafique2024machine, mutambik2024efficient} or timing analyses for fixed control channels \cite{wang2003robust, houmansadr2009rainbow}, existing implementations are evaluated within offline simulations or static fabrics. This paper focuses on engineering and empirically validating a combined network-telemetry, packet-timing, and hardware-attestation defense framework on an active ROS 2 fleet under multi-layer adversarial exploits.

Architectural defense design must examine the underlying middleware. ROS 2 relies on DDS as its core middleware layer, utilizing the Real-Time Publish-Subscribe (RTPS) protocol over UDP/IP for discovery and message orchestration \cite{macenski2022robot}. In standard tactical edge deployments, DDS instances operate via unauthenticated multi-cast or uni-cast discovery routines. Because standard profiles do not leverage encryption, critical geometric velocity vectors (\texttt{geometry\_msgs/msg/Twist}) and sensory telemetry streams are transmitted as plaintext. This absence of identity verification exposes the system to interception, malicious replay, and unauthenticated command-injection from any adversary within the wireless broadcast domain \cite{deng2022security, diluoffo2019credential}. While the DDS-Security specification introduces cryptographic plugins, the compute and packet-size overhead generated by these mechanisms degrade the tight feedback loops required by resource-constrained single-board computers (SBCs).

To support resilient architectures for tactical setups, three core technical security capabilities are examined. First, timing-based flow watermarking provides a non-cryptographic packet-layer verification mechanism that embeds a distribution-based statistical timing signature into discrete frame departure intervals, allowing a receiver to evaluate sample kurtosis ($K$) over a rolling window to isolate user-space scheduling perturbations introduced by unauthorized MITM relays \cite{wang2003robust, houmansadr2009rainbow}. Second, Security Information and Event Management (SIEM) systems bridge fleet-wide observability by aggregating multi-source audit logs via lightweight telemetry shippers (Packetbeat, Metricbeat, and Heartbeat) to track network and host states without exhausting constrained compute cycles. Third, a hardware-bound root of trust leverages a Trusted Platform Module (TPM 2.0) coprocessor to securely isolate Endorsement Keys (EK) and execute cryptographic hash measurements during boot to enforce node attestation \cite{bravi2025embrave}. 

Rather than deploying these mechanisms in isolation, our integrated approach leverages their concurrent capabilities to enforce a specialized Zero Trust architecture across the tactical network. Continuous session verification is enforced by the packet-level IPD watermarking to detect anomalous traffic flows, context-aware observability is achieved via centralized SIEM threshold monitoring of multi-source telemetry, and explicit asset identity is validated by the TPM hardware root of trust to systematically prevent rogue device enrollment and identity masquerading within the tactical fleet.


\section{Threat Model and Approach}

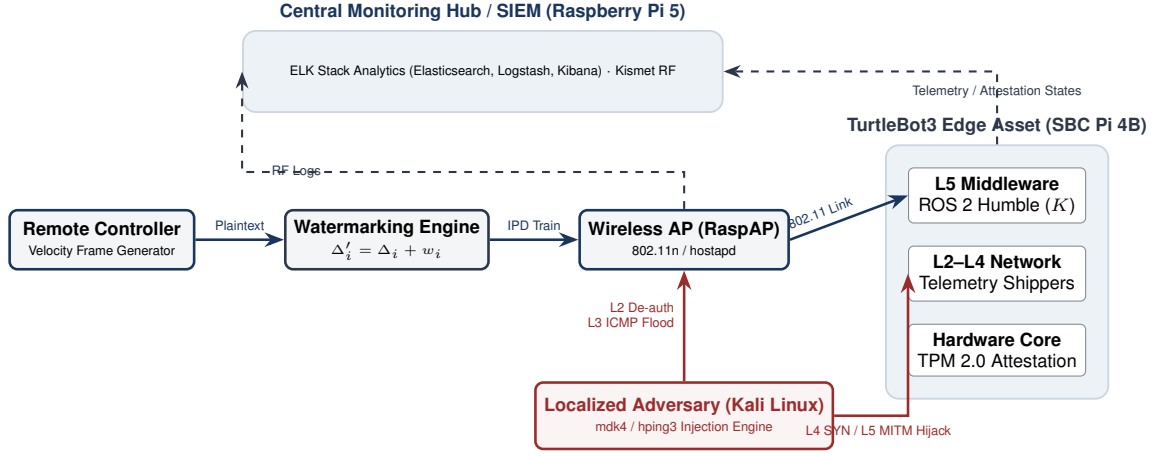
\begin{figure*}[!t]
\centering
\resizebox{0.85\textwidth}{!}{%
\begin{tikzpicture}[
    font=\sffamily\scriptsize,
    node distance=0.8cm and 1.4cm,
    >=Stealth,
    core/.style={rectangle, draw=navy, fill=navy!5, minimum width=2.5cm, minimum height=0.8cm, align=center, rounded corners=3pt, line width=1pt},
    engine/.style={rectangle, draw=slate, fill=slate!5, minimum width=2.5cm, minimum height=0.8cm, align=center, rounded corners=3pt, line width=1pt},
    threat/.style={rectangle, draw=crimson, fill=crimson!5, minimum width=3.2cm, minimum height=0.9cm, align=center, rounded corners=3pt, line width=1pt, text=crimson},
    enclosure/.style={rectangle, draw=bordergray, fill=ice, rounded corners=5pt, line width=0.7pt},
    innerblock/.style={rectangle, draw=slate!60, fill=white, minimum width=2.4cm, minimum height=0.6cm, align=center, rounded corners=2pt, font=\sffamily\scriptsize}
]

    \node[core] (controller) {\textbf{Remote Controller} \\ \tiny{Velocity Frame Generator}};
    
    \node[engine, right=1.2cm of controller] (watermarker) {\textbf{Watermarking Engine} \\ \tiny{$\Delta'_i = \Delta_i + w_i$}};
    
    \node[core, right=1.2cm of watermarker] (gateway) {\textbf{Wireless AP (RaspAP)} \\ \tiny{802.11n / hostapd}};

    \node[innerblock, right=1.6cm of gateway, yshift=0.6cm] (applayer) {\textbf{L5 Middleware} \\ ROS 2 Humble ($K$)};
    \node[innerblock, below=0.3cm of applayer] (netlayer) {\textbf{L2--L4 Network} \\ Telemetry Shippers};
    \node[innerblock, below=0.3cm of netlayer] (hwlayer) {\textbf{Hardware Core} \\ TPM 2.0 Attestation};
    
    \begin{scope}[on background layer]
        \node[enclosure, fit={(applayer) (netlayer) (hwlayer)}, inner sep=0.3cm, label={[slate]above:\textbf{TurtleBot3 Edge Asset (SBC Pi 4B)}}] (robot) {};
    \end{scope}

    \node[threat, below=1.5cm of gateway] (adversary) {\textbf{Localized Adversary (Kali Linux)} \\ \tiny{mdk4 / hping3 Injection Engine}};

    \node[enclosure, minimum width=6.5cm, minimum height=1.1cm, align=center, above=1.3cm of watermarker, xshift=1.3cm, label={[navy]above:\textbf{Central Monitoring Hub / SIEM (Raspberry Pi 5)}}] (siem) {\tiny{ELK Stack Analytics (Elasticsearch, Logstash, Kibana) $\cdot$ Kismet RF}};

    \draw[->, line width=1pt, navy] (controller) -- node[above, font=\sffamily\tiny] {Plaintext} (watermarker);
    \draw[->, line width=1pt, navy] (watermarker) -- node[above, font=\sffamily\tiny] {IPD Train} (gateway);
    \draw[->, line width=1pt, navy] (gateway.east) -- node[above, sloped, pos=0.3, font=\sffamily\tiny] {802.11 Link} (applayer.west);

    \draw[->, line width=1pt, crimson] (adversary.north) -- node[left, pos=0.6, font=\sffamily\tiny, align=right] {L2 De-auth \\ L3 ICMP Flood} (gateway.south);
    \draw[->, line width=1pt, crimson] (adversary.east) -| node[below, pos=0.3, font=\sffamily\tiny] {L4 SYN / L5 MITM Hijack} (netlayer.west);

    \draw[->, line width=0.8pt, dashed, slate] (gateway.north) -- ++(0,0.5cm) -| node[left, pos=0.4, font=\sffamily\tiny] {RF Logs} (siem.west);
    \draw[->, line width=0.8pt, dashed, slate] (robot.north) |- node[above, pos=0.25, font=\sffamily\tiny] {Telemetry / Attestation States} (siem.east);

\end{tikzpicture}
}
\caption{Multi-layered Zero-Trust architecture pipeline detailing the operational control loop flow, adversarial attack vectors (OSI Layers 2--5), and the security capabilities.}
\label{fig:zero_trust_pipeline}
\end{figure*}

\begin{table*}[!htbp]
\centering
\caption{Integrated Zero-Trust Security Matrix Mapping}
\label{tab:zero_trust_mapping}
\begin{adjustbox}{width=\textwidth}
\begin{tabular}{lllll}
\toprule
\textbf{Zero-Trust Principle} & \textbf{Defensive Layer} & \textbf{Security Capability} & \textbf{Targeted OSI Layer} & \textbf{Mitigated Threats} \\
\midrule
Asset Identity & Hardware Identity & TPM 2.0 Coprocessor / EK Attestation & Hardware Initialization & Node Spoofing, Cloning, Rogue Enrollment \\
Observability & Network Telemetry & ELK Stack SIEM \& Packetbeat / Kismet Shippers & OSI Layer 2 \& 3 & 802.11 De-authentication, ICMP Flood \\
Observability & Host State Auditing & Metricbeat \& Heartbeat Audit Logs & OSI Layer 4 & TCP SYN Resource / Socket Exhaustion \\
Continuous Verification & Packet Timing & Inter-Packet Delay (IPD) Watermarking ($K$) & OSI Layer 5 & MITM Content Spoofing, Command Injection \\
\bottomrule
\end{tabular}
\end{adjustbox}
\end{table*}

This section expands on the adversarial threat model, details the proposed multi-layer defense approach, and summarizes how it aligns with Zero-Trust. Important to note here, that this work adopt a red-blue team (often referred to as purple team) perspective to validate risks and assess defenses. 

\subsection{Threat Model/Red Team}
The red offensive capability is based on a threat model with an on-path or localized adversary operating within the physical and wireless broadcast proximity of the tactical fleet with the objective of disrupting the timing bounds of the real-time feedback loop or hijacking the control signals themselves. By inducing arbitrary packet transit latencies or directly overwriting topic payloads, the adversary causes critical path divergence, spatial boundary violation, or vehicle immobilization, undermining operational availability and integrity introducing significant public-safety risks~\cite{yaacoub2020security, alrefaei2022survey}.

The adversary is assumed to be equipped with standard commercial wireless hardware and open-source penetration testing tools. The adversary exploits the absence of authentication and verification across the local network and middleware layers \cite{deng2022security, diluoffo2019credential}. The threat landscape is categorized into four distinct vectors across the open systems interconnection (OSI) model:
\begin{enumerate}
    \item \textit{Layer 2 Availability Degradation:} The adversary initiates localized, high-rate 802.11 de-authentication floods via \textit{mdk4}, targeting unauthenticated Wi-Fi management frames to force constant endpoint disassociation.
    \item \textit{Layer 3 Network-Layer Saturation:} The adversary executes ICMP echo-request floods via \textit{hping3}, depleting the network interface card (NIC) buffers and compute capacity of the edge nodes.
    \item \textit{Layer 4 Transport Exhaustion:} The adversary targets embedded operational remote management (e.g., SSH on Port 22) with a high-velocity TCP SYN flood, exhausting kernel socket connection allocation tables.
    \item \textit{Layer 5 Middleware Integrity Violation:} The adversary manipulates neighbor discovery tables via localized ARP cache poisoning to insert an active Man-in-the-Middle (MITM) relay. The adversary then captures plaintext ROS2 geometry messages, replacing legitimate command vectors with forged velocity instructions.
\end{enumerate}

\subsection{The Multilayer Integrated Defense Approach/Blue Team}
To achieve comprehensive defensive coverage, the framework integrates three security capabilities that operate at the hardware initialization, network flow telemetry, and packet transmission stages.

\subsubsection{Centralized SIEM Telemetry}
Real-time infrastructure visibility and security analytics are achieved by a centralized Security Operations Center (SOC) running an ELK Stack (comprising Elasticsearch, Logstash, and Kibana) hosted on an embedded proxy platform (Raspberry Pi 5) \cite{elasticELKStack}. The centralized SOC ingests multi-source data streams from lightweight, low-overhead log forwarders (data shippers) embedded on each robotic asset. Real-time frames and Layer 3 protocols are parsed via \textit{Packetbeat}, while host-level socket allocations, system memory state, and CPU utilization parameters are monitored via \textit{Metricbeat} \cite{elasticBeatsPlatform}. Network interface accessibility states are evaluated continuously via \textit{Heartbeat} ping sequences. This is augmented by another monitoring node executing \textit{Kismet} to continuously sample the ambient 802.11 spectrum, generating immediate alerts upon the appearance of unauthenticated broadcast disassociation frame vectors \cite{kismet}. Further details are included in \ref{sec:exp}.

\subsubsection{Packet-Layer Inter-Packet Delay (IPD) Watermarking}
To identify on-path data manipulation without introducing cryptographic overheads, a timing-based flow watermarking scheme is formalized. Let $t_i$ denote the nominal transmission timestamp of the $i$-th velocity command packet emitted by the controller. The baseline IPD is represented by $\Delta_i = t_i - t_{i-1}$. The watermarking engine modulates packet release schedules by embedding a stationary pseudo-random timing offset sampled from a zero-mean Gaussian distribution, yielding a modified delay interval $\Delta'_i$:
\begin{equation}
    \Delta'_i = \Delta_i + w_i, \quad w_i \sim \mathcal{N}(\mu, \sigma^2)
\end{equation}
here, let $\mu = 0$ and $\sigma = 0.02$\,s. Upon packet arrival at the robotic node, the timing signature is evaluated over a rolling observation window of size $N$ (in our case $N = 250$). The detector computes the sample Kurtosis ($K$) of the incoming delay offsets:
\begin{equation}
    K = \frac{\frac{1}{N} \sum_{i=1}^{N} (\Delta'_i - \bar{\Delta}')^4}{\left( \frac{1}{N} \sum_{i=1}^{N} (\Delta'_i - \bar{\Delta}')^2 \right)^2}
\end{equation}
Under uncompromised conditions, the incoming packet sequence reflects the Gaussian variance, maintaining a steady sample kurtosis bound of $2.8 \le K \le 3.2$. Because an active MITM relay introduces queuing dynamics and arbitrary OS task-scheduling jitter to intercept and parse topic payloads, an adversary inevitably corrupts this timing pattern. This perturbation flattens the distribution or drives heavy-tailed outliers, pushing $K$ outside the expected baseline ($K < 2.0$ or $K > 4.5$) and triggering an integrity alert \cite{wang2003robust, houmansadr2009rainbow}.

\subsubsection{Hardware-Based Root of Trust}
Static endpoint identity is enforced by equipping each node with a hardware-based Trusted Platform Module (TPM 2.0) cryptographic coprocessor \cite{bravi2025embrave}. During the initialization phase, the asset must complete an automated challenge-response attestation routine. The platform verifies the unique Endorsement Key (EK) digital signature and evaluates integrity check hashes across system registers. Rogue devices or cloned assets lacking a verified, pre-mission registered hardware module are systematically isolated and blocked from entering the fleet topology.

\subsection{Zero-Trust Strategic Mapping}
The individual hardware, network, and packet components are mapped directly to core Zero-Trust principles in Table~\ref{tab:zero_trust_mapping}.

\begin{figure}[htbp]
    \centering
    \includegraphics[width=1\linewidth]{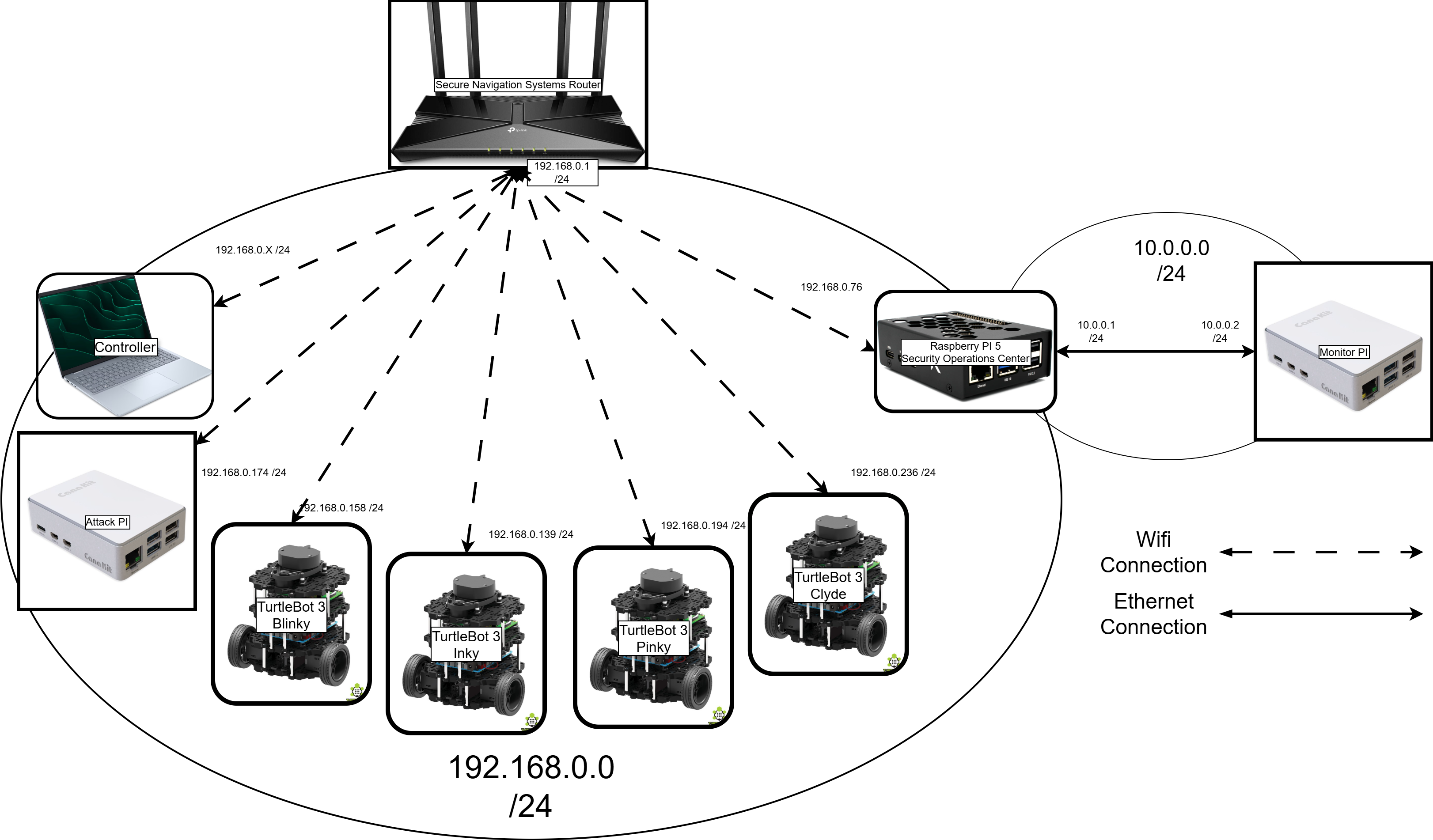}
    \caption{Robotic Testbed Connectivity Setup}
    \label{fig:network_setup}
    \vspace{-15pt}
\end{figure}

\section{Experimental Configuration and Empirical Results}
\label{sec:exp}
This section details the physical hardware infrastructure, software middleware components, and the corresponding per-layer empirical evaluations across the four multi-layer adversarial attack scenarios.
\begin{figure}[htbp]
    \centering
    \includegraphics[width=1\linewidth]{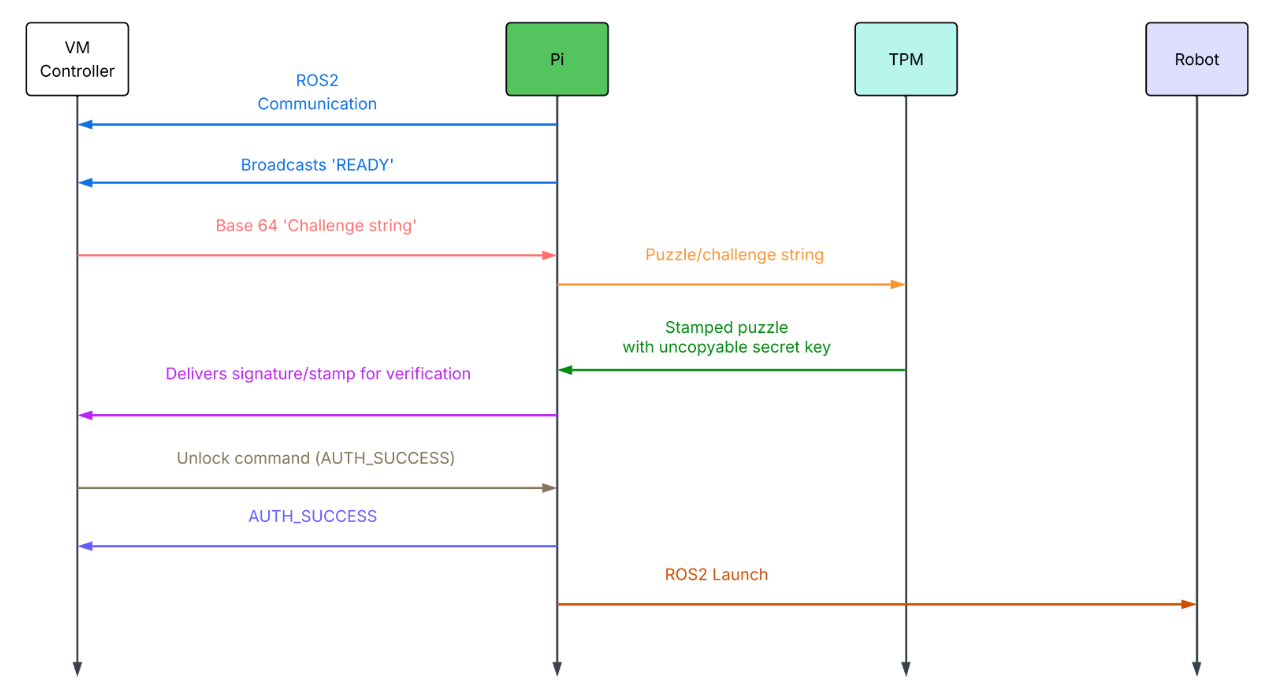}
    \caption{TPM and robot initialization.}
    \label{fig:tpm}
    \vspace{-10pt}
\end{figure}

\begin{figure}[htbp]
    \centering
    \includegraphics[width=1\linewidth]{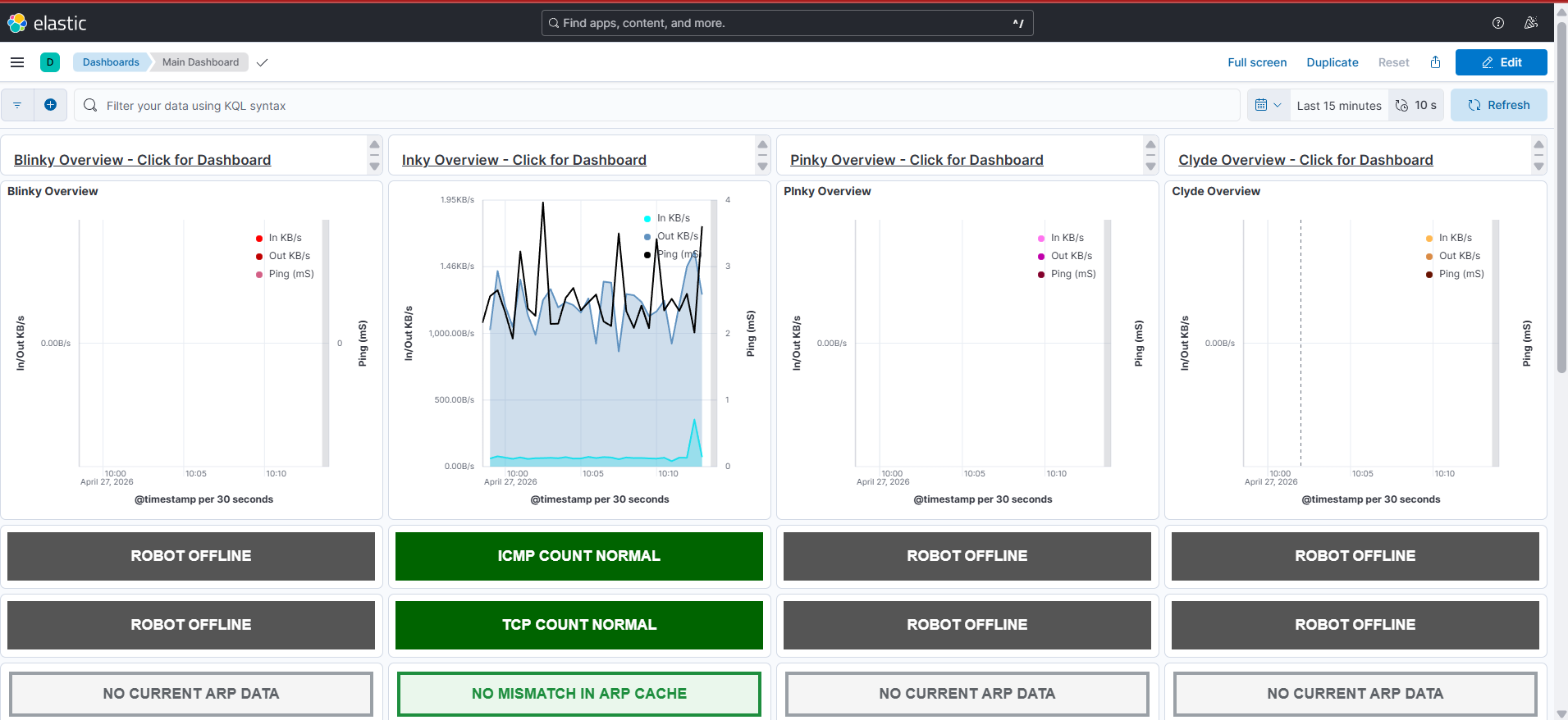}
    \caption{SOC dashboard - monitoring screen.}
    \label{fig:dashboardMain}
    \vspace{-10pt}
\end{figure}

\begin{figure}[htbp]
    \centering
    \includegraphics[width=1\linewidth]{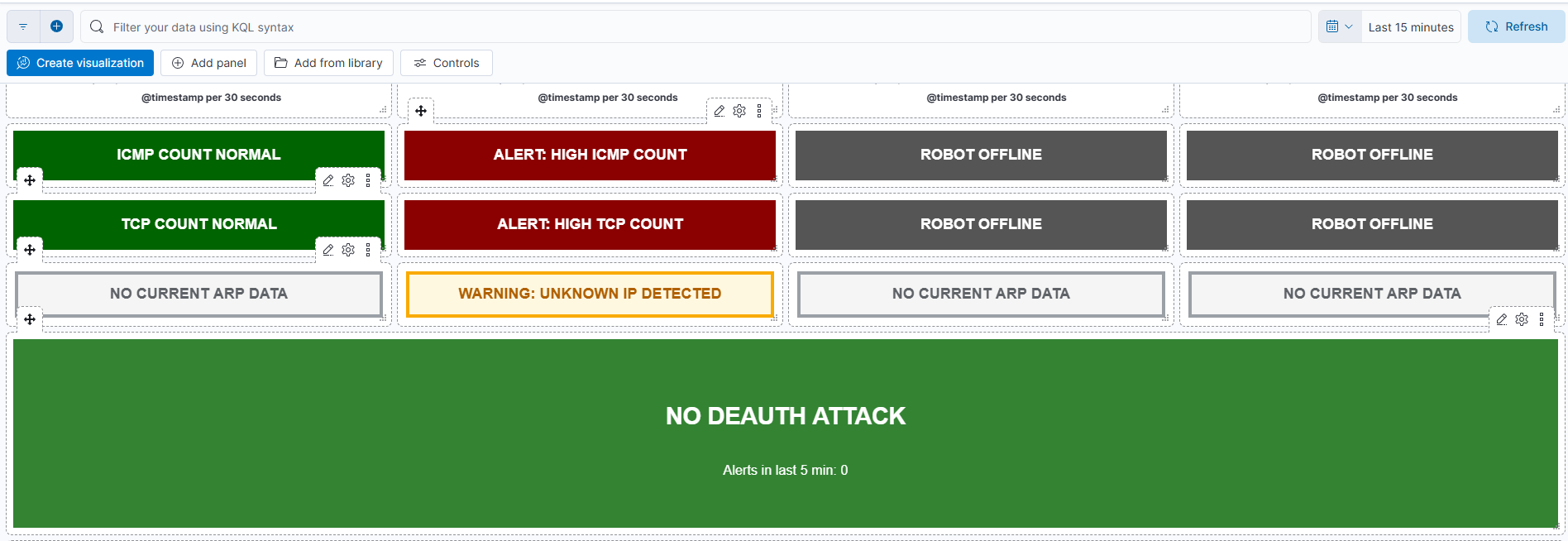}
    \caption{SOC dashboard under ICMP and TCP SYN floods.}
    \label{fig:dashboardattack}
    \vspace{-10pt}
\end{figure}

\subsubsection{Hardware Testbed, Software Ecosystem, and Baseline Instrumentation}
The experimental validation platform is composed of an autonomous mobile ground vehicle (TurtleBot3 Burger platform) communicating with a remote centralized command station over an IEEE 802.11n wireless local area network, as illustrated in the network topology map in Fig.~\ref{fig:network_setup}. The ground asset relies on an embedded single-board computer (SBC), a Raspberry Pi 4B with 4\,GB RAM running an ARM64 Linux kernel, to execute navigation, coordinate state transformations, and process network packets. The routing infrastructure is hosted on an equivalent Raspberry Pi 4 executing an isolated wireless access point configuration via \textit{hostapd} and \textit{RaspAP}. The distributed robotic control loops operate on the Robot Operating System (ROS 2 Humble Hawksbill flavor) running over the default eProsima Fast DDS middleware layer. Operational control loops are supported through transmitting periodic geometric velocity vectors (\texttt{geometry\_msgs/msg/Twist}) from the controller topic publisher to the vehicle subscriber at a deterministic frequency of 50\,Hz.
\begin{table}[htbp]
\centering
\scriptsize{
\caption{Adversarial commands and parameters}
\label{tab:attack_commands}
\begin{tabular}{lp{6.2cm}}
\toprule
\textbf{Parameter} & \textbf{Configuration Value / Terminal Executable} \\
\midrule
\textbf{Target OS} & Ubuntu 22.04 LTS (Linux Kernel 5.15.0-91-generic) \\
\textbf{Middleware} & ROS 2 Humble Hawksbill (eProsima Fast DDS) \\
\textbf{OSI Layer 2} & \texttt{sudo mdk4 wlan1mon d -B [Robot\_MAC] -g} \\
\textbf{OSI Layer 3} & \texttt{sudo hping3 --icmp --flood [Robot\_IP]} \\
\textbf{OSI Layer 4} & \texttt{sudo hping3 -S -p 22 --flood [Robot\_IP]} \\
\textbf{OSI Layer 5} & \texttt{sudo arpspoof -i wlan0 -t [Robot\_IP] [AP\_IP]} \\
                     & \texttt{sudo arpspoof -i wlan0 -t [AP\_IP] [Robot\_IP]} \\
\bottomrule
\end{tabular}
}
\vspace{-10pt}
\end{table}
Real-time continuous observability is driven by an enterprise-grade Security Information and Event Management (SIEM) architecture as detailed previously. Network endpoint reachability is verified via \textit{Heartbeat} pings, creating the unified baseline visibility displayed in the primary SOC dashboard in Fig.~\ref{fig:dashboardMain}. Spectrum monitoring is performed with \textit{Kismet}. At the physical tier, the mobile asset's embedded computing core is coupled with an active hardware-based TPM coprocessor to execute pre-mission (EK) identity attestation, introducing an initial startup delay of 314\,ms prior to fleet network association, following the sequence shown in Fig.~\ref{fig:tpm}.

\subsubsection{Offensive Subsystem Configuration and Attack Scenario 1: Link-Layer Flooding}
The offensive subsystem consists of a Raspberry Pi 4B running Kali Linux, paired with a TP-Link AC600 dual-band USB wireless adapter providing monitor mode and stable packet injection capabilities. The commands and parameters selected for the red team are summarized in Table.~\ref{tab:attack_commands}. In the first scenario, the adversary executes a continuous 802.11 de-authentication flood using \textit{mdk4}, spoofing the MAC address of the legitimate access point and broadcasting de-authentication frames directed at the robot's wireless interface. This exploits the unauthenticated nature of 802.11 management frames under WPA2. Empirically, the de-authentication flood instantly disconnected all devices associated with the \textit{RaspAP} gateway. This immediate link disconnection severed the control feedback loop, preventing the robot from receiving further motion instructions and causing it to veer off its pre-programmed elliptical trajectory. Because the underlying connection was totally dropped, no standard network metrics could be captured on the interface. However, the out-of-band monitoring node executing \textit{Kismet} successfully intercepted the unauthorized management frames and generated a real-time high-priority alert.
\begin{figure}[htbp]
    \centering
    \includegraphics[width=0.85\linewidth]{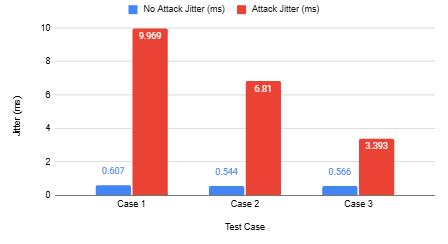}
    \caption{Bar Graph Comparing Jitter in Scenarios With and Without ICMP Attacks}
    \label{fig:jitterlayer3}
    \vspace{-10pt}
\end{figure}

\begin{figure}[htbp]
    \centering
    \includegraphics[width=0.85\linewidth]{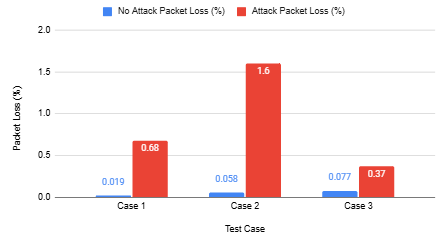}
    \caption{Bar Graph Comparing Packet Loss in Scenarios With and Without ICMP Attacks}
    \label{fig:packetlosslayer3}
    \vspace{-10pt}
\end{figure}
\subsubsection{Attack Scenario 2: Network-Layer Saturation (Layer 3 ICMP Flood)}
The offensive node executed a volumetric Layer 3 ICMP flood directed at the robot's static IP address using \textit{hping3}, transmitting echo-request packets at a maximum send rate to saturate the robot's network interface card (NIC) buffers. To quantify the network degradation, link parameters were measured using \textit{iperf3} (executing \texttt{iperf3 -c [IP] -u -b 2M} on the client and \texttt{iperf3 -s} on the server). Under nominal conditions, the baseline packet jitter ranged between 0.5\,ms and 0.6\,ms, with a baseline packet loss rate bounded between 0.02\% and 0.08\%. As shown in the comparative empirical distributions in Fig.~\ref{fig:jitterlayer3} and Fig.~\ref{fig:packetlosslayer3}, active ICMP flooding forced packet jitter to escalate sharply to a range of 3.0\,ms to 10.0\,ms, while packet loss degraded heavily to a range of 0.4\% to 1.6\%. This network-layer saturation directly fractured feedback synchronization, causing the robot to perform elongated turns, delay steering adjustments, and exhibit inconsistent physical travel patterns that deviated heavily from the pre-defined path. The attack was successfully isolated by the SIEM threshold monitor via a specialized detection script that flagged the anomalous volumetric surge in incoming ICMP echo-requests, as captured on the active threat dashboard view in Fig.~\ref{fig:dashboardattack}.

\begin{figure}[htbp]
    \centering
    \includegraphics[width=0.85\linewidth]{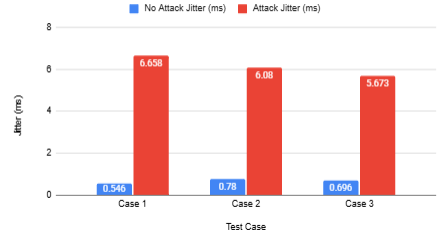}
    \caption{Bar Graph Comparing Jitter in Scenarios With and Without TCP SYN Flood Attacks}
    \label{fig:jitterlayer4}
    \vspace{-10pt}
\end{figure}

\begin{figure}[htbp]
    \centering
    \includegraphics[width=0.85\linewidth]{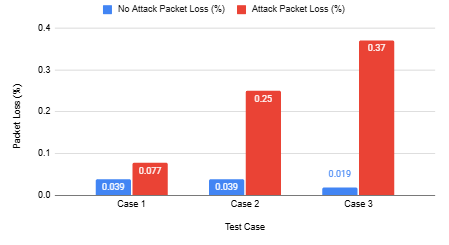}
    \caption{Bar Graph Comparing Packet Loss in Scenarios With and Without TCP SYN Flood Attacks}
    \label{fig:packetlosslayer4 }
    \vspace{-10pt}
\end{figure}
\subsubsection{Attack Scenario 3: Transport-Layer Exhaustion (Layer 4 TCP SYN Flood)}
The adversary initiated a Layer 4 TCP SYN flood using \textit{hping3} directed at Port 22 (SSH service) on the robot's IP address, transmitting high-velocity connection initialization requests to exhaust the host's kernel socket allocation tables. Empirical evaluations showed severe transport degradation, which is quantified across trials in Fig.~\ref{fig:jitterlayer4} and Fig.~\ref{fig:packetlosslayer4 }. Active flooding forced packet jitter to jump from a baseline range of 0.5\,ms--0.8\,ms up to an elevated range of 5.6\,ms--6.6\,ms. Concurrently, packet loss increased from a nominal baseline of 0.02\%--0.04\% to an active attack range of 0.08\%--0.4\%. Physically, this resource starvation caused the robot to experience intermittent control-loop disassociation, manifesting as frequent, complete stops where the asset immobilized mid-transit due to delayed velocity frame processing. The \textit{Metricbeat} telemetry shipper forwarded host socket states to the ELK engine, where a Kibana detection script successfully flagged the exploit via a static volume threshold tuned to anomalous SYN state transitions, aligning with the real-time alerts shown in Fig.~\ref{fig:dashboardattack}.
\begin{figure}[htbp]
    \centering
    \includegraphics[width=0.85\linewidth]{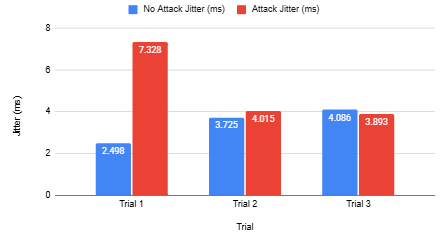}
    \caption{Comparing jitter with and without MITM attacks.}
    \label{fig:jitterlayer5}
    \vspace{-10pt}
\end{figure}

\begin{figure}[htbp]
    \centering
    \includegraphics[width=0.85\linewidth]{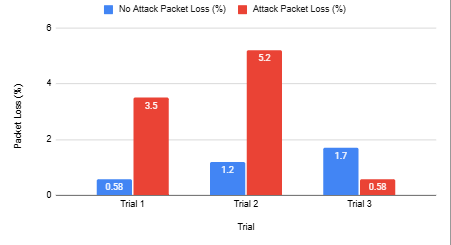}
    \caption{Scenario-based packet loss comparison with and without MITM attacks.}
    \label{fig:packetlosslayer5}
    \vspace{-10pt}
\end{figure}
\begin{figure}[htbp]
    \centering
    \includegraphics[width=0.75\linewidth]{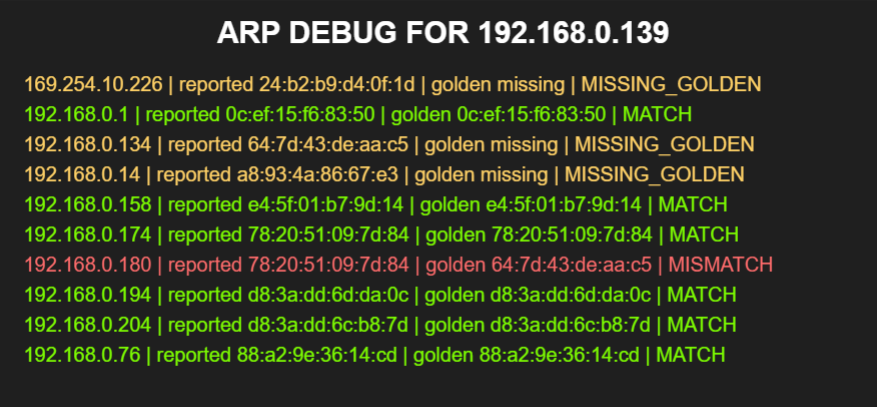}
    \caption{ARP table change detection on Kibana dashboard. }
    \label{fig:layer5_arp}
    \vspace{-10pt}
\end{figure}

\subsubsection{Attack Scenario 4: Middleware Integrity Violation (Layer 5 MITM Command Injection)}
The final scenario evaluated a sophisticated on-path Man-in-the-Middle (MITM) attack designed to intercept and modify ROS 2 velocity commands without altering volumetric thresholds. The attacker executed localized ARP cache poisoning to redirect the traffic flowing between the controller and the robot through the malicious workstation. IP forwarding was enabled via \texttt{sysctl}, and kernel-level interception was established via \texttt{iptables} NFQUEUE rules. A custom Python script dequeued incoming packets, extracted the unencrypted DDS middleware payloads, injected forged linear and angular speed values via a manual mapping interface, and re-injected the tampered frames into the network stream. 

As demonstrated by the empirical data in Fig.~\ref{fig:jitterlayer5} and Fig.~\ref{fig:packetlosslayer5}, standard network metrics proved entirely blind to this exploit: packet jitter during the active MITM relay remained nearly identical to the baseline jitter, while packet loss exhibited high variance, swinging negligibly both slightly above and below the baseline boundaries. 
\begin{figure}[htbp]
    \centering
    \includegraphics[width=0.85\linewidth]{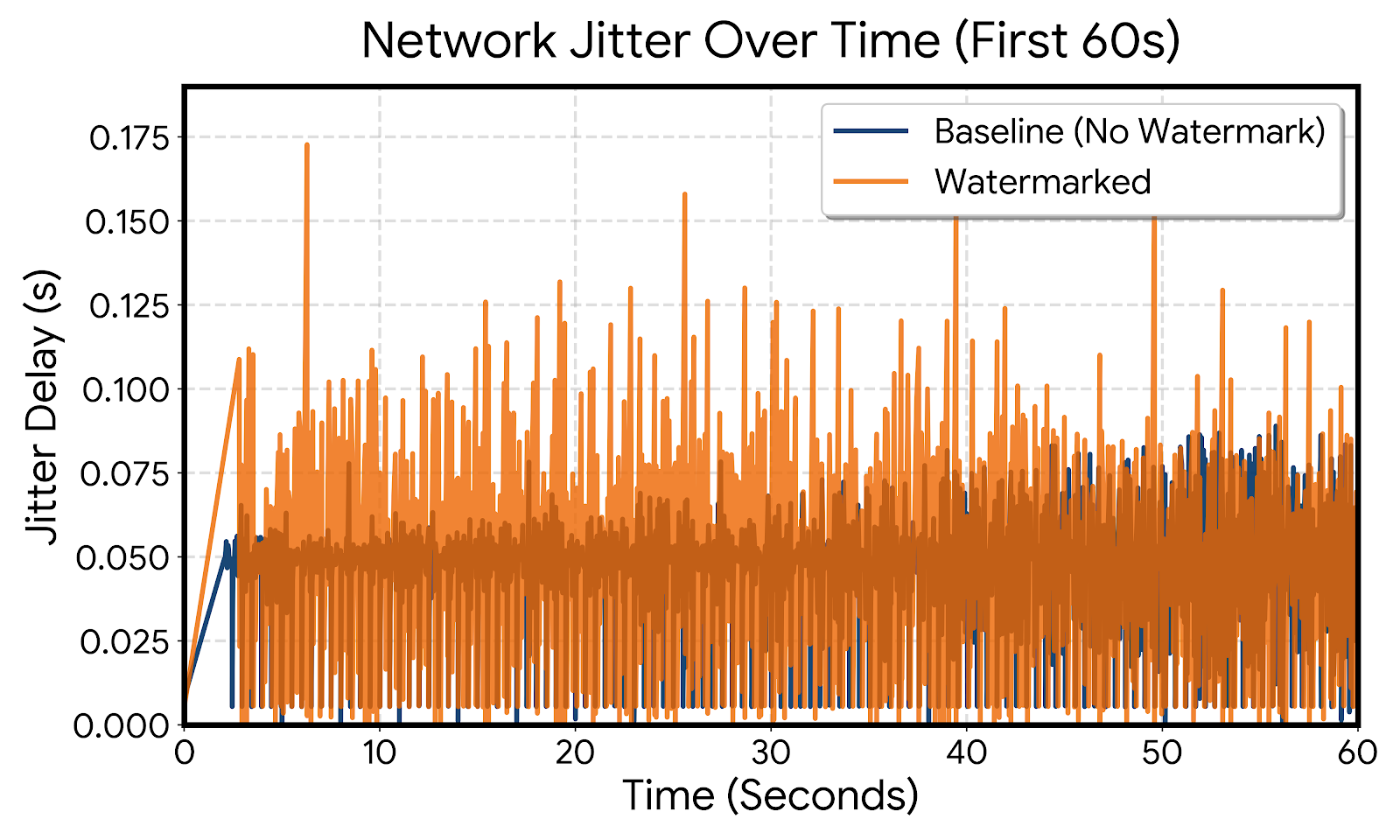}
    \caption{Jitter time-series for watermarked and non-watermarked packets.}
    \label{fig:ipd-time}
    \vspace{-10pt}
\end{figure}

\begin{figure}[htbp]
    \centering
    \includegraphics[width=0.85\linewidth]{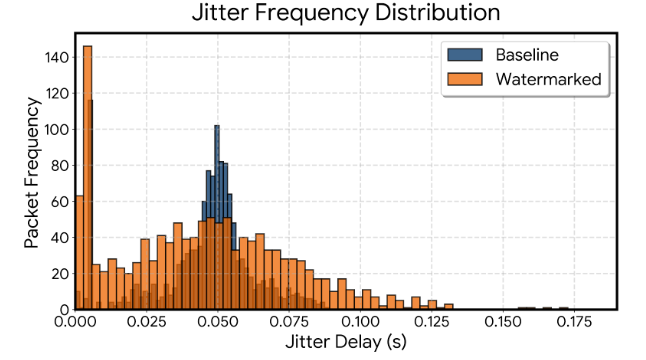}
    \caption{Jitter histogram for watermarked and non-watermarked packets.}
    \label{fig:ipd-hist}
    \vspace{-10pt}
\end{figure}

To detect this stealthy compromise, the IPD watermarking scheme was deployed at the controller, injecting a zero-mean Gaussian timing offset ($w_i \sim \mathcal{N}(0, 0.02^2)$\,s) into the packet release schedule, introducing a visible timing modulation over time as shown in Fig.~\ref{fig:ipd-time} and distributed statistically in Fig.~\ref{fig:ipd-hist}. While the nominal baseline watermark maintained a steady sample kurtosis bound of $2.8 \le K \le 3.2$, the user-space queuing latency and task-scheduling jitter introduced by the attacker's Python relay altered the watermark distribution, driving a sharp kurtosis spike ($K < 2.0$ or $K > 4.5$). By compiling timing metrics over a rolling 5-second observation window, the IPD watermarking detector achieved a complete 20/20 success rate in identifying active injection attacks, while maintaining a 19/20 success rate under legitimate watermarked conditions. Concurrently, the SIEM dashboard's ARP consistency monitor crossed referenced the network state against a static reference database, generating a 10/10 perfect detection rate of the unauthorized IP-to-MAC mapping within seconds of the initial poisoning phase, as documented by the data dashboard log in Fig.~\ref{fig:layer5_arp}.

\section{Conclusion and Future Work}

This paper presented and empirically evaluated a layered Zero-Trust cybersecurity framework for hardware-constrained wireless robotic networks. By integrating a TPM 2.0 hardware root of trust, centralized multi-source SIEM telemetry (ELK Stack and Kismet), and a non-cryptographic Inter-Packet Delay (IPD) timing watermark, the architecture establishes multi-tier resilience without application-layer cryptographic overhead. Testbed evaluations under multi-layer exploits (OSI Layers 2--5) demonstrated that while volume-based filters isolate brute denial-of-service floods, tracking the statistical sample kurtosis ($K$) of the IPD watermark exposes stealthy Layer 5 Man-in-the-Middle command injections with complete detection accuracy. Future work will replace manual node configurations with an automated, zero-touch deployment pipeline for rapid tactical scaling in public safety missions, and adapt the real-time IPD and TPM loops to support heterogeneous fleets, specifically targeting the strict flight dynamics of unmanned aerial vehicles (UAVs).

\section*{Acknowledgment}
Gemini and Claude AI tools were employed in refining the
manuscript technical narrative.


\bibliographystyle{IEEEtran}
\bibliography{references}

@IEEEtranBSTCTL{IEEEexample:BSTcontrol,
  CTLdash_repeated_names = "no"
}

@article{alcorn2026darrms,
  title={DARRMS--An Efficient Algorithm for Dynamic Attention Radius in Resource-Constrained Multi-Agent Systems},
  author={Alcorn, Benjamin and Hammad, Eman},
  journal={arXiv preprint arXiv:2606.12614},
  year={2026}
}

@inproceedings{alcorn2025situational,
  title={Situational awareness for safe and robust multi-agent interactions under uncertainty},
  author={Alcorn, Benjamin and Hammad, Eman},
  booktitle={2025 IEEE World Forum on Public Safety Technology (WF-PST)},
  pages={36--41},
  year={2025},
  organization={IEEE}
}

@article{farraj2025reading,
  title={Reading lips: An analytical framework for adversarial passive detection of wireless traffic in IoT ecosystems},
  author={Farraj, Abdallah and Hammad, Eman},
  journal={IEEE Access},
  year={2025},
  publisher={IEEE}
}

@article{farraj2024physical,
  title={A physical-layer security cooperative framework for mitigating interference and eavesdropping attacks in Internet of Things environments},
  author={Farraj, Abdallah and Hammad, Eman},
  journal={Sensors},
  volume={24},
  number={16},
  pages={5171},
  year={2024},
  publisher={MDPI}
}

@article{yaacoub2020security,
  title={Security analysis of drones systems: Attacks, limitations, and recommendations},
  author={Yaacoub, Jean-Paul and Noura, Hassan and Salman, Ola and Chehab, Ali},
  journal={Internet of Things},
  volume={11},
  pages={100218},
  year={2020},
  publisher={Elsevier}
}

@inproceedings{alrefaei2022survey,
  title={A survey on the jamming and spoofing attacks on the unmanned aerial vehicle networks},
  author={Alrefaei, Faisal and Alzahrani, Abdullah and Song, Houbing and Alrefaei, Salma},
  booktitle={2022 IEEE International IOT, Electronics and Mechatronics Conference (IEMTRONICS)},
  pages={1--7},
  year={2022},
  organization={IEEE}
}

@article{zeng2026gnss,
  title={GNSS Jamming and Spoofing Threats in UAV Navigation: Countermeasure Status and Challenges},
  author={Zeng, Yejia and Lu, Zukun and Zhao, Xiaoyu and Xiao, Zhu and Ni, Shaojie and Han, Zhu and Li, Keqin},
  journal={IEEE Communications Surveys \& Tutorials},
  year={2026},
  publisher={IEEE}
}

@article{macenski2022robot,
  title={Robot operating system 2: Design, architecture, and uses in the wild},
  author={Macenski, Steven and Foote, Tully and Gerkey, Brian and Lalancette, Chris and Woodall, William},
  journal={Science robotics},
  volume={7},
  number={66},
  pages={eabm6074},
  year={2022},
  publisher={American Association for the Advancement of Science}
}

@inproceedings{deng2022security,
  title={On the (in) security of secure ros2},
  author={Deng, Gelei and Xu, Guowen and Zhou, Yuan and Zhang, Tianwei and Liu, Yang},
  booktitle={Proceedings of the 2022 ACM SIGSAC Conference on Computer and Communications Security},
  pages={739--753},
  year={2022}
}

@article{diluoffo2019credential,
  title={Credential masquerading and openssl spy: Exploring ros 2 using dds security},
  author={DiLuoffo, Vincenzo and Michalson, William R and Sunar, Berk},
  journal={arXiv preprint arXiv:1904.09179},
  year={2019}
}

@article{rafique2024machine,
  title={Machine learning and deep learning techniques for internet of things network anomaly detection—current research trends},
  author={Rafique, Saida Hafsa and Abdallah, Amira and Musa, Nura Shifa and Murugan, Thangavel},
  journal={Sensors},
  volume={24},
  number={6},
  pages={1968},
  year={2024},
  publisher={MDPI}
}

@misc{elasticELKStack,
  author = {{Elasticsearch B.V.}},
  title = {The {Elastic Stack (ELK Stack)}: {Elasticsearch, Logstash, and Kibana}},
  howpublished = {\url{https://www.elastic.co/elastic-stack}},
  year = {2026},
  note = {Accessed: May 2026}
}

@misc{elasticBeatsPlatform,
  author = {{Elasticsearch B.V.}},
  title = {{Beats: Lightweight Data Shippers for Elasticsearch (Packetbeat, Metricbeat, Heartbeat)}},
  howpublished = {\url{https://www.elastic.co/beats}},
  year = {2026},
  note = {Accessed: May 2026}
}

@inproceedings{kismet,
  title={Detecting rogue access points using kismet},
  author={Thejdeep, G and Sagar, B Shiva and Siddartha, LK and Chandavarkar, BR},
  booktitle={2015 International Conference on Communications and Signal Processing (ICCSP)},
  pages={0172--0175},
  year={2015},
  organization={IEEE}
}

@article{mutambik2024efficient,
  title={An efficient flow-based anomaly detection system for enhanced security in IoT networks},
  author={Mutambik, Ibrahim},
  journal={Sensors},
  volume={24},
  number={22},
  pages={7408},
  year={2024},
  publisher={MDPI}
}

@inproceedings{wang2003robust,
  title={Robust correlation of encrypted attack traffic through stepping stones by manipulation of interpacket delays},
  author={Wang, Xinyuan and Reeves, Douglas S},
  booktitle={Proceedings of the 10th ACM conference on Computer and communications security},
  pages={20--29},
  year={2003}
}

@inproceedings{houmansadr2009rainbow,
  title={RAINBOW: A robust and invisible non-blind watermark for network flows.},
  author={Houmansadr, Amir and Kiyavash, Negar and Borisov, Nikita},
  booktitle={NDSS},
  volume={47},
  pages={406--422},
  year={2009}
}

@article{bravi2025embrave,
  title={EMBRAVE: EMBedded remote attestation and verification framework},
  author={Bravi, Enrico and Claudio, Alessio and Lioy, Antonio and Vesco, Andrea},
  journal={Sensors},
  volume={25},
  number={17},
  pages={5514},
  year={2025},
  publisher={MDPI}
}
\end{document}